\documentclass[preprintnumbers, floatfix, preprintnumbers, letterpaper, twocolumn, superscriptaddress,nofootinbib]{revtex4}
\usepackage{graphicx}
\usepackage{microtype}
\usepackage{amsmath}
\usepackage{amssymb}
\usepackage{subfigure}
\usepackage{hyperref}
\usepackage{url}
\usepackage{xcolor}
\usepackage{color}
\usepackage{mathrsfs}
\usepackage{calrsfs}
\usepackage{amsfonts}
\usepackage{tabularx}
\usepackage{eucal}
\usepackage{latexsym}
\usepackage{ragged2e}
\usepackage{epsfig}
\usepackage{textcomp}
\usepackage{float}

\usepackage{caption}
\DeclareCaptionJustification{justified}{\leftskip=0pt \rightskip=0pt \parfillskip=0pt plus 1fil}
\definecolor{vividviolet}{rgb}{0.62, 0.0, 1.0}
\definecolor{amaranth}{rgb}{0.9, 0.17, 0.31}
\definecolor{palatinateblue}{rgb}{0.15, 0.23, 0.89}
\definecolor{brightpink}{rgb}{1.0, 0.0, 0.5}
\definecolor{cornflowerblue}{rgb}{0.39, 0.58, 0.93}
\definecolor{deepcarminepink}{rgb}{0.94, 0.19, 0.22}
\definecolor{radicalred}{rgb}{1.0, 0.21, 0.37}

\hypersetup{ linktoc=all,
	colorlinks, linkcolor={palatinateblue},
	citecolor={brightpink}, urlcolor={amaranth}
}

\graphicspath{{Images/}}

\renewcommand{\d}[1]{\ensuremath{\operatorname{d}\!{#1}}}

\def\sideremark#1{\ifvmode\leavevmode\fi\vadjust{\vbox to0pt{\vss
			\hbox to 0pt{\hskip\hsize\hskip1em
				\vbox{\hsize1.3cm\tiny\raggedright\pretolerance10000
					\noindent #1\hfill}\hss}\vbox to8pt{\vfil}\vss}}}%
\def\beq{\begin{equation}}
\def\eeq{\end{equation}}

\begin{document}
\title{An Effective Sign Switching Dark Energy:\\ Lotka-Volterra Model of Two Interacting Fluids}

\author{Yen Chin \surname{Ong}}
\email{ycong@yzu.edu.cn}
\affiliation{Center for Gravitation and Cosmology, College of Physical Science and Technology, Yangzhou University, \\180 Siwangting Road, Yangzhou City, Jiangsu Province  225002, China}
\affiliation{Shanghai Frontier Science Center for Gravitational Wave Detection, School of Aeronautics and Astronautics, Shanghai Jiao Tong University, Shanghai 200240, China}

\begin{abstract}
One of the recent attempts to address the Hubble and $S_8$ tensions is to consider the Universe started out not as a de Sitter-like spacetime, but rather anti-de Sitter-like. That is, the Universe underwent an ``AdS-to-dS'' transition at some point. We study the possibility that there are two dark energy fluids, one of which gave rise to the anti-de Sitter-like early Universe. The interaction is modeled by the Lotka-Volterra equations, commonly used in population biology. We consider ``competition'' models that are further classified as ``unfair competition'' and ``fair competition''. The former involves a quintessence in competition with a phantom, and the second involves two phantom fluids. Surprisingly, even in the latter scenario it is possible for the overall dark energy to cross the phantom divide. The latter model also allows a \emph{constant} $w$ ``AdS-to-dS'' transition, thus evading the theorem that such a dark energy must possess a singular equation of state. We also consider a ``conversion'' model in which a phantom fluid still manages to achieve ``AdS-to-dS'' transition even if it is being converted into a negative energy density quintessence. In these models, the energy density of the late time effective dark energy is related to the coefficient of the quadratic self-interaction term of the fluids, which is analogous to the resource capacity in population biology. 
\end{abstract} 

\maketitle

\section{Introduction: Cosmology with Sign Switching Dark Energy}

The Hubble tension \cite{1607.05617,1907.10625,2008.11284,2103.01183} and the $S_8$ tension \cite{1409.2769,2005.03751,2008.11285} in cosmology continue to be highly debated \cite{1907.07569, 1908.03663, 1911.06456, 2002.11707,2203.10558,2203.06142,2202.11852,2208.14435,2209.11476,2209.14330}. The former is the mismatch between the locally measured expansion rate and the inferred rate via the cosmic microwave background (CMB), while the latter concerns the measurement of galaxy clusters on a scale of 8 $h^{-1}$Mpc, which revealed that matter has not clumped as much as expected assuming the concordance $\Lambda$CDM cosmology and its parameters constraints given by the CMB data. (For a review on these issues, as well as other challenges facing $\Lambda$CDM cosmology, see \cite{2105.05208}. See \cite{1205.3421} for an introduction to various dark energy scenarios). If these effects are real, they could be due to modified gravity or other new physics \cite{2103.02117, 2201.09848}.

Note that in the $\Lambda$CDM model, the Hubble parameter as a function of the redshift $H(z)$, is specified by two constant fitting parameters \cite{2206.11447}: $(H_0, \Omega_m)$ or equivalently $(A,B)$ as follows:
\begin{flalign}
H(z)^2&=H_0^2\left[1-\Omega_m+\Omega_m(1+z)^3\right]\\ \notag
&=A+B(1+z)^3,
\end{flalign}
where $A$ is the term associated with dark energy. The aforementioned tensions could mean that $\Lambda$CDM is not correct and thus the ``constant'' fitting parameters could evolve with redshift (or equivalently, with time). In \cite{2206.11447} it was noted that using constraints from $H(z)$, observed data exhibit an increasing $\Omega_m$ (decreasing $H_0$) trend with increasing bin redshift, and yields a `pile up' around $A=0$ or $\Omega=1$. If $\Omega$ can increase beyond unity, this amounts to the dark energy density switching sign. (See also \cite{2211.02129,2212.00238}.)

Indeed, one of the possible ways to ameliorate these tensions is to consider the possibility that the Universe was originally more anti-de Sitter-(AdS)-like than de Sitter-(dS)-like \cite{1808.06623,1811.03505,1907.07953,1912.08751,2001.02451,2006.16291,2008.10237,2102.05701,2107.13286,2108.09239,2112.10641,2202.12214,2203.13037,2205.09311,2208.05583,2211.05742,2212.00050}. That is, the physics of dark energy (DE) might be more complicated than we initially expected\footnote{Indeed, such a possibility was already considered from other perspectives before the Hubble tension became a serious issue \cite{0307185,0403104,1105.0078,1105.2636,1807.01570}. See also the recent work \cite{2211.12611}.}. The main idea is to reduce the tension between the higher value of $H_0$ obtained from CMB with the lower value obtained by local measurements, by changing cosmological models either at the recombination epoch or at late time \cite{2002.11707}. 
For example, a phantom energy at late time would accelerate cosmic expansion faster than a cosmological constant would. In addition, Baryon Oscillation Spectroscopic Survey (BOSS) of SDSS-III probing the Ly$\alpha$ forest of quasars also indicates preference for a positive dark energy density at late time but a negative one at early time \cite{1404.1801}. Future observations such as SKA \cite{1811.02743}, BINGO \cite{2107.01633,2107.01639}, and Euclid \cite{1606.00180} could potentially further constrain this possibility. Another observational support for negative energy density comes from Pantheon+ data of high redshift supernovae \cite{2301.12725}.

Such a possibility can be realized by simply promoting the cosmological constant $\Lambda$ to\footnote{Here $\text{sgn}$ is the sign function (i.e., it is $+1$ for positive argument and $-1$ for negative argument).} $\Lambda_s := \Lambda_{s0} \text{sgn}[z_\dagger-z]$, where $\Lambda_{s0}$ is the present value of the cosmological constant, and $z_\dagger$ is the value of redshift at which the sign switching abruptly happened \cite{2108.09239}. See also \cite{2104.02623}. 
Another model considers a ``graduated dark energy'' with energy density and pressure satisfying $\rho + p \propto \rho^\lambda$, which provides a continuous transition (controlled by the parameter $\lambda$) from AdS-like to dS-like Universe \cite{1912.08751}. Other options include the possibility that the dark energy sector could consist of a negative cosmological constant \emph{and} a phantom dark energy \cite{1907.07953} or a quintessence \cite{2112.10641}. (See, however, \cite{2202.03906}.) A different approach based on fractal modification to the entropy via a running Barrow index \cite{2004.09444} could also give rise to an effective sign-changing dark energy  \cite{2205.09311}. 

In this work, we shall consider what happens if instead of a scalar field on top of a negative cosmological constant, we have either (1) a quintessence with negative energy density, which competes with a phantom dark energy with a positive energy density\footnote{The idea that different dark energy components might interact with each other is not new. For example, models in which a quintessence interacts with a Chaplygin gas was considered in \cite{1104.3983}, in an attempt to explain the coincidence problem \cite{1410.2509}.}, or (2) a phantom with a negative energy density that competes with another phantom with a positive energy density, or (3) a phantom with positive energy density being converted into a negative energy density quintessence. We shall refer to these scenarios, respectively, as ``unfair competition'', ``fair competition'', and ``conversion'' models.
This is inspired by the interacting models between dark matter (DM) and dark energy \cite{0707.2089,1412.4091,1603.08299,2209.09685,2209.14816,2301.06097}, as well as from biological species interactions. In fact, the connection between these two subjects has been noticed in the literature. For example, in \cite{1306.1037}  Perez et al., as well as Aydiner  in \cite{1610.07338}, pointed out that the DM-DE interaction can be re-written as the Lotka-Volterra equation, which is commonly used in population biology to model the interactions between various species. In cosmological contexts, Lotka-Volterra equation was also studied in \cite{1603.02267,1603.07620}. In the population model, it is of course required that the numbers of the species involved are non-negative, whereas in our model, the corresponding quantities are the energy densities of the DE fluids, which can be negative by assumption. It should furthermore be mentioned that if one considers non-minimal interactions between DE and DM to reduce the Hubble tension, the $S_8$ tension would in turn be exacerbated, hence we consider the alternative of non-minimal interaction between two DE fields to get one effective DE that exhibits sign-switching energy density, which in principle can address both tensions simultaneously. Our work is only meant to be an illustration of concept with the simplest models. Some assumptions would need to be relaxed or improved before a more realistic model can be used for data fitting the actual universe.

We will work out  the conditions on the interactions between two dark energy fluids (``DE-DE interaction'') in order to obtain a late time accelerated expansion with a very small but positive energy density. In our models, unlike the single fluid models in the literature, neither fluid exhibits any singular behavior in their equation of state, although if the phantom divide is crossed, the combined effective dark energy typically does exhibit such a singular behavior during the AdS-to-dS transition. Remarkably, we found that in the fair competition model, it is possible for the effective dark energy to cross the phantom divide despite both component fluids satisfy $w<-1$.
In addition, in this scenario there are evolutions that allow AdS-to-dS transition without crossing the phantom divide, which therefore is free of singular behavior in its equation of state.
Finally, it is often said that the fact that the dark energy density is extremely small is ``unnatural''. We shall see that in the two-fluid model, this value is related to the coefficient of the quadratic self-interaction term of the fluids, which mathematically plays the same role as the resource capacity of a biological population.

\section{The Unfair Competition Model}

In  \cite{1306.1037}  and \cite{1610.07338}, the authors considered a model in which dark energy is being converted into dark matter via 
\begin{equation}\label{1}
\begin{cases}
\dot{\rho}_{\textnormal{\tiny \textsc{DE}}} + 3H(\rho_\textnormal{\tiny \textsc{DE}}+p_\textnormal{\tiny \textsc{DE}})= -\gamma \rho_\textnormal{\tiny \textsc{DE}} \rho_\textnormal{\tiny \textsc{DM}},\\
\\
\dot{\rho}_\textnormal{\tiny \textsc{DM}} + 3H(\rho_\textnormal{\tiny \textsc{DM}}+p_\textnormal{\tiny \textsc{DM}})= \gamma \rho_\textnormal{\tiny \textsc{DE}} \rho_\textnormal{\tiny \textsc{DM}},
\end{cases}
\end{equation}
where $\gamma>0$. Likewise, dark matter can be converted into dark energy with $\gamma<0$. See \cite{1902.09684} for generalizations.

The equations of state considered in \cite{1610.07338} are $w_{\textnormal{\tiny \textsc{DM}}}\geqslant 0$ and $w_{\textnormal{\tiny \textsc{DE}}}<-1$, i.e., dark matter is a ``normal matter'' while dark energy is a phantom fluid. Thus, one can define two positive quantities: 
\begin{equation}\label{r}
\begin{cases}
R_1:=-3H(1+w_{\textnormal{\tiny \textsc{DE}}}) > 0,\\
R_2:=3H(1+w_{\textnormal{\tiny \textsc{DM}}}) > 0.
\end{cases}
\end{equation}
Furthermore, assuming that the Hubble parameter is slowly varying (so that $R_1$ and $R_2$ are both approximately constant) and upon introducing\footnote{One can check that in the units in which the speed of light $c=1$, $R_1$ and $R_2$ have physical dimension $[\text{time}]^{-1}$, while $\gamma$ has dimension $[\text{density}\cdot \text{time}]^{-1}$; while $x_1$ and $x_2$ are dimensionless.} $x_1:=\gamma\rho_{\textnormal{\tiny \textsc{DE}}}/r_2$ and  $x_2:=\gamma\rho_{\textnormal{\tiny \textsc{DM}}}/r_1$, Eq.(\ref{1}) can be re-written as
\begin{equation}\label{LV1}
\begin{cases}
\dfrac{\d x_1}{\d t}=R_1x_1-R_1x_1x_2,\\
\\
\dfrac{\d x_2}{\d t}=R_2x_1x_2-R_2x_2,
\end{cases}
\end{equation}
which is explicitly a Lotka-Volterra equation that describes a predator-prey dynamic with $x_1$ being the ``prey'' and $x_2$ the ``predator''. The system thus displays an oscillatory behavior. Such an interacting model could potentially help to resolve the coincidence problem.

In our case, we have two interacting dark energy fluids, which we will denote by $\Lambda_1 < 0$ and $\Lambda_2 > 0$ (despite the notation, they are not constant; the notation is meant to remind us that they are mimicking cosmological constants). 
Their energy densities would be denoted by $\rho_{\Lambda_1}$ and $\rho_{\Lambda_2}$, respectively.
The transition from an early time AdS-like Universe to a late time dS-like spacetime thus amounts to  $\Lambda_1$ becoming subdominant to $\Lambda_2$.
Let us first consider an unrealistic model (to be improved upon later) that is analagous to Eq.(\ref{1}), so we can point out the differences:
\begin{equation}\label{2}
\begin{cases}
\dot{\rho}_{\Lambda_1} + 3H(\rho_{\Lambda_1}+p_{\Lambda_1})=-\gamma\rho_{\Lambda_1}\rho_{\Lambda_2},\\
\\
\dot{\rho}_{\Lambda_2} + 3H(\rho_{\Lambda_2}+p_{\Lambda_2})=\gamma\rho_{\Lambda_1}\rho_{\Lambda_2},\\
\end{cases}
\end{equation}
where $\gamma > 0$. As before, we assume that the late time dark energy is a phantom, thus $w_{\Lambda_2} < -1$. On the other hand\footnote{In the cosmological case, AdS spacetime has $w=-1$, which is the same as dS spactime. Unlike dS spacetime, however, in AdS spacetime the negative cosmological constant corresponds to a negative energy density but with a positive pressure. Cosmological evolution with negative energy densities was previously studied in details in \cite{2205.01619}.}, $w_{\Lambda_1}>-1$ but with $\rho_{\Lambda_1}<0$. This is justified by \cite{2201.11623,2202.01202}, in which it was argued that solving both the Hubble and $S_8$ tensions require the overall effective dark energy to cross the phantom divide (if we assume that Newton's gravitational constant is not varying). See also \cite{2005.12587}. This is also similar to the model in \cite{1912.08751}. 

Thus we define, analogous to Eq.(\ref{r}), two parameters
\begin{equation}\label{r2}
\begin{cases}
r_1:=3H(1+w_{\Lambda_1}) > 0,\\
r_2:=-3H(1+w_{\Lambda_2}) > 0.
\end{cases}
\end{equation}
We assume that both $w_{\Lambda_1}$ and $w_{\Lambda_2}$ are constant.
Upon introducing the dimensionless quantities $x:=\gamma \rho_{\Lambda_1}/r_2$ and $y:=\gamma \rho_{\Lambda_2}/r_1$, we obtain the Lotka-Volterra equation of the form
\begin{equation}\label{1b}
\begin{cases}
\dfrac{\d x}{\d t}=-r_1x-r_1xy,\\
\\
\dfrac{\d y}{\d t}=r_2xy+r_2y, 
\end{cases}
\end{equation}
whose some signs are different from that of interacting DM-DE model in Eq.(\ref{LV1}), and with $x<0, y>0$. This system does not oscillate, but rather there is an attractor $x \to 0^-$ and $y \to \infty$.
Note that in contrast to the DM-DE interaction model, it is not quite right to say that $\Lambda_1$ is being ``converted'' into $\Lambda_2$ here (a true conversion model will be studied in Sec.(\ref{convert})), since $x<0$ implies that the interaction term is \emph{negative} for both $x$ and $y$. In other words, the two fluids are competing, but unlike two competing species whose birth rates are both positive, $x$ is itself
diminishing exponentially due to the ``death rate'' term $-r_1x$. Hence the name ``unfair competition''.
It is clear that the phantom fluid thus dominates over the quintessence. 
That is, $\rho_{\Lambda_1}$ is asymptotically vanishing while $\rho_{\Lambda_2}$ becomes large (and eventually diverges) at late time. This is not a desired property since we know from observation that dark energy density is only of the order of $10^{-30} \text{g/cm}^3$.

We can improve upon this model by modifying the $\d y/\d t$ term so that

\begin{equation}\label{model2}
\begin{cases}
\dfrac{\d x}{\d t}=-r_1x-r_1xy,\\
\\
\dfrac{\d y}{\d t}= r_2 xy + r_2y\left(1-\dfrac{y}{K}\right),
\end{cases}
\end{equation}
where $K>0$ is a constant. The attractor is then $x \to 0^-$ and $y \to K$, so we can prescribe to $K$ the observed value.
At this point this seems rather ad hoc, but later on we will give it a physical interpretation.

\begin{figure}[h!]
\begin{center}
\includegraphics[width=0.45\textwidth]{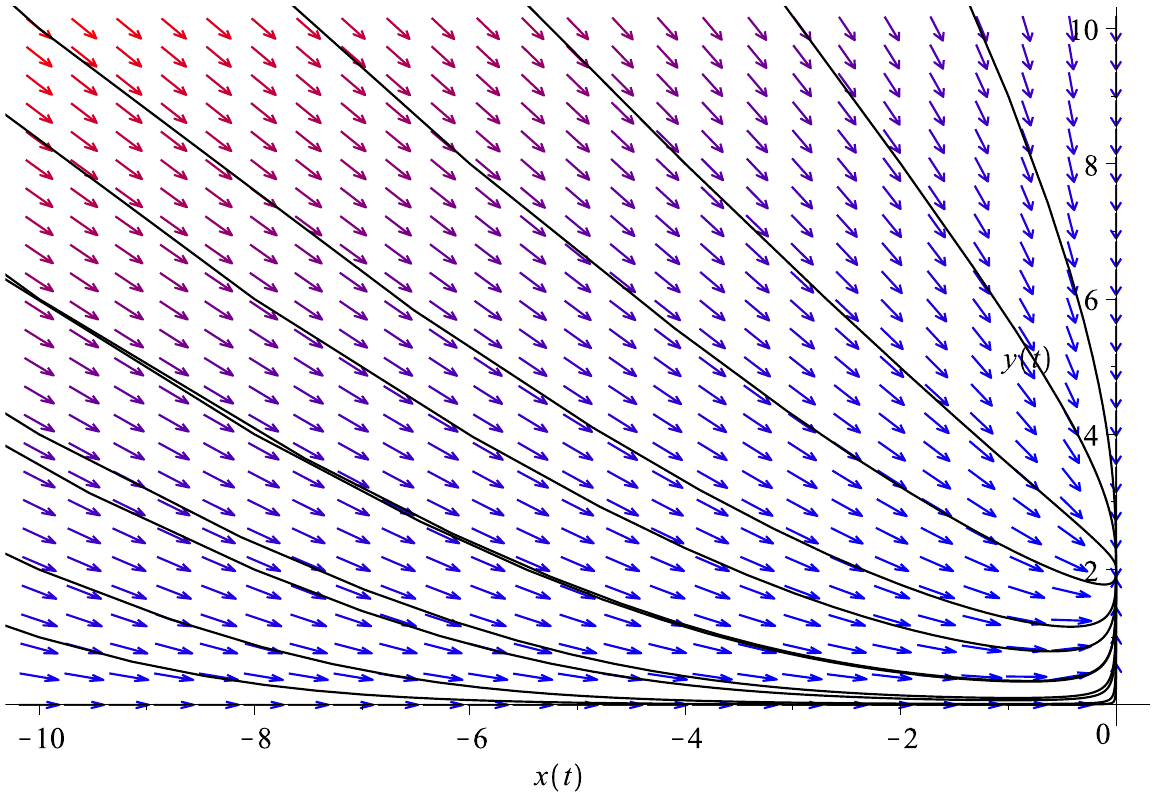}
\caption{The phase diagram of the two dark energy fluid model in Eq.(\ref{model2}). Here we use $r_1=r_2=1$ and $K=2$. The system exhibits a fixed point at $(x,y)=(0,K)$.}
\label{figflow}
\end{center}
\end{figure}

The analogy to population biology can also be made here: with $x$ small at late time, in the absence of $y/K$ term, what we have is analogous to an exponential growth population ${\d y}/{\d t}= r_2y$, whereas 
Eq.(\ref{model2}) corresponds to the logistic model with a resource capacity (or ``carrying capacity'') $K$, so that the actual population size cannot diverge but rather asymptotes to a constant value. 
Note that since $x \to 0^-$ anyway, there is no good reason to add the resource term to $\d x/\d t$ in this model. A typical phase diagram is given in Fig.(\ref{figflow}).

\begin{figure}
\begin{center}
\includegraphics[width=0.45\textwidth]{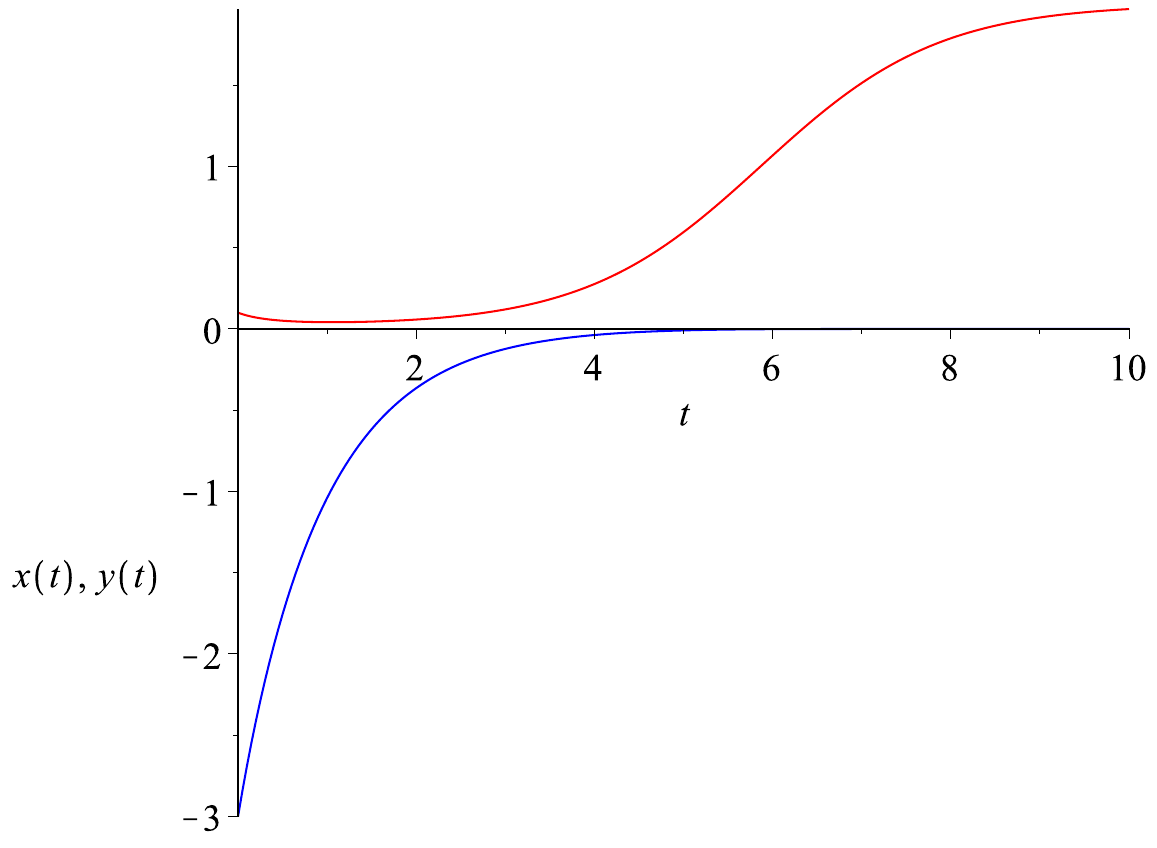}
\includegraphics[width=0.45\textwidth]{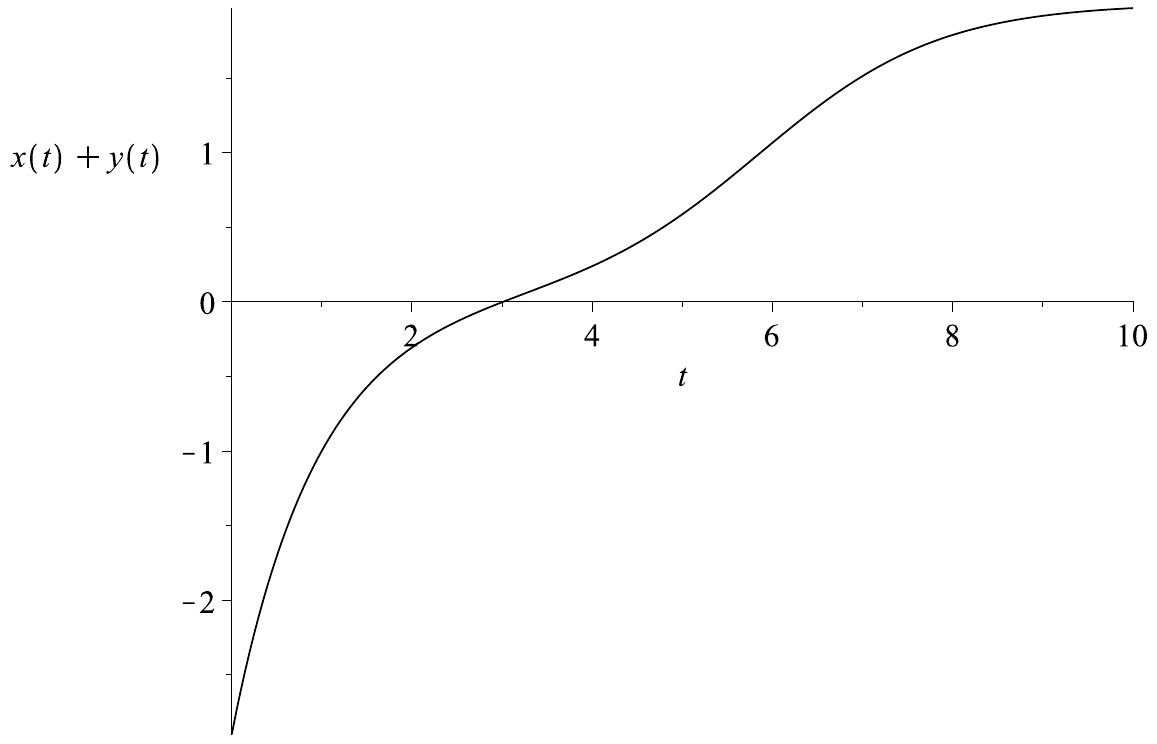}
\caption{\textbf{Top:} The evolution of $x(t)$ (bottom curve in blue) and $y(t)$ (top curve in red) with the initial condition set to be $x(0)=-3,y(0)=0.1$, and with $r_1=1=r_2,K=2$. \textbf{Bottom:} The evolution of $x(t)+y(t)$, which is essentially the re-scaled version of  $\rho_{\Lambda_1}(t)+\rho_{\Lambda_2}(t)$ (in this example, they are in fact equal).}
\label{ex1}
\end{center}
\end{figure}

Note that the effective dark energy density is the sum $\rho_{\Lambda_1}+\rho_{\Lambda_2}$. 
Thus, in order that this quantity starts out negative, we need the initial condition to satisfy $y(0)<|x(0)|$. Then, since the attractor is $(x,y)=(0,K)$ with $K>0$, it follows that by continuity $x(t)+y(t)$ must cross-over to $y>0$ at some point. The exact profile of $\rho_{\Lambda_1}(t)+\rho_{\Lambda_2}(t)$ or the re-scaled equation $x(t)+y(t)$ would depend on the initial conditions, but the transition from an overall negative energy density to a positive one can be smoother than the model in \cite{1912.08751}. One example is given in Fig.(\ref{ex1}). 

\begin{figure}
\begin{center}
\includegraphics[width=0.46\textwidth]{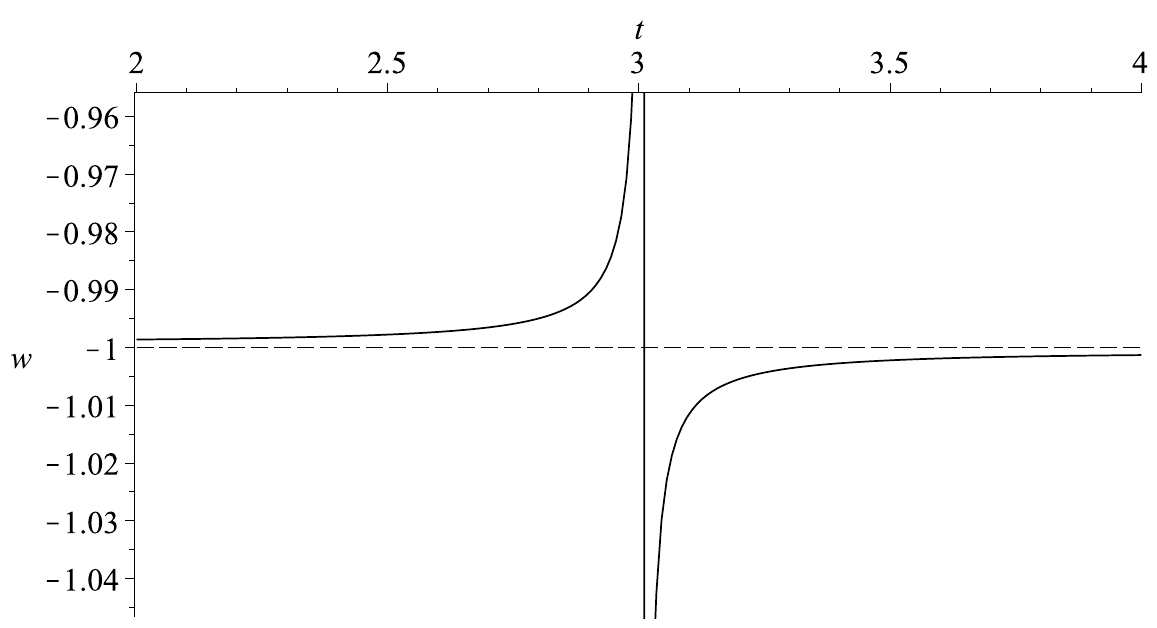}
\caption{The evolution of the overall effective $w$ of the example in Fig.(\ref{ex1}) with $w_{\Lambda_2}=-0.999$ and $w_{\Lambda_1}=-1.001$.}
\label{w1}
\end{center}
\end{figure}

If we know what kind of profile is desired from observational constraints, this would in turn provides us a mean to choose the coefficients $r_1$ and $r_2$, as well as the initial conditions of the Lotka-Volterra equation. We also note that the equation of states of both dark energy components are never singular, though the combined effective dark energy density has to pass through $\rho_{\Lambda_1}(t)+\rho_{\Lambda_2}(t)=0$, and the effective equation of state \emph{is} singular at that point \cite{2203.04167}. This can be seen as follows: if we consider the combined fluid to still satisfy the equation of state of the form $p=w\rho$, where $\rho:=\rho_{\Lambda_1}+\rho_{\Lambda_2}$ and $p:=p_{\Lambda_1}+p_{\Lambda_2}$, then
\begin{equation}
\rho=\rho_{\Lambda_1}+\rho_{\Lambda_2}=w^{-1}(p_{\Lambda_1}+p_{\Lambda_2}),
\end{equation}
which implies that the effective varying $w$ is the weighted average:
\begin{equation}\label{w}
w=\frac{w_{\Lambda_1}\rho_{\Lambda_1}+w_{\Lambda_2}\rho_{\Lambda_2}}{\rho_{\Lambda_1}+\rho_{\Lambda_2}}=\frac{w_1r_2x_1+w_2r_1x_2}{r_2x_1+r_1x_2}.
\end{equation}
Therefore, evidently $w \to \pm \infty$ when the denominator is zero. Again, the situation is similar to the ``graduated dark energy'' model of \cite{1912.08751}, though in that case there is only one dark energy fluid, whose equation of state becomes singular. An example of the evolution of $w$ is shown in Fig.(\ref{w1}).
Note that the values of $w_{\Lambda_1}$ and $w_{\Lambda_1}$ are not freely prescribed since they are constrained by Eq.(\ref{r2}), in the sense that once we fix $r_1$ and $r_2$, the relations between $w_{\Lambda_1}$ and $w_{\Lambda_2}$ are also determined.
In our example, choosing $r_1=r_2=1$ implies $w_{\Lambda_1}+w_{\Lambda_2}=-2$.

We remark that the sign of $x(t)+y(t)$ is not necessarily the same as the sign of $\rho_{\Lambda_1}(t)+\rho_{\Lambda_2}(t)$. In fact,
\begin{equation}
x(t)+y(t)=\frac{\gamma}{r_1r_2}\left(r_1\rho_{\Lambda_1}(t)+r_2\rho_{\Lambda_2}(t)\right).
\end{equation}
Thus the sign of $x(t)+y(t)$  is the same as the sign of $r_1\rho_{\Lambda_1}(t)+r_2\rho_{\Lambda_2}(t)$.
For simplicity our example deals with $r_1=r_2$, and the two expressions do have the same sign, and furthermore $w$ is singular when $x+y=0$ or equivalently at $r_1\rho_{\Lambda_1}+r_2\rho_{\Lambda_2}=0$.

\section{The Fair Competition Model}

\begin{figure}[h!]
\begin{center}
\includegraphics[width=0.45\textwidth]{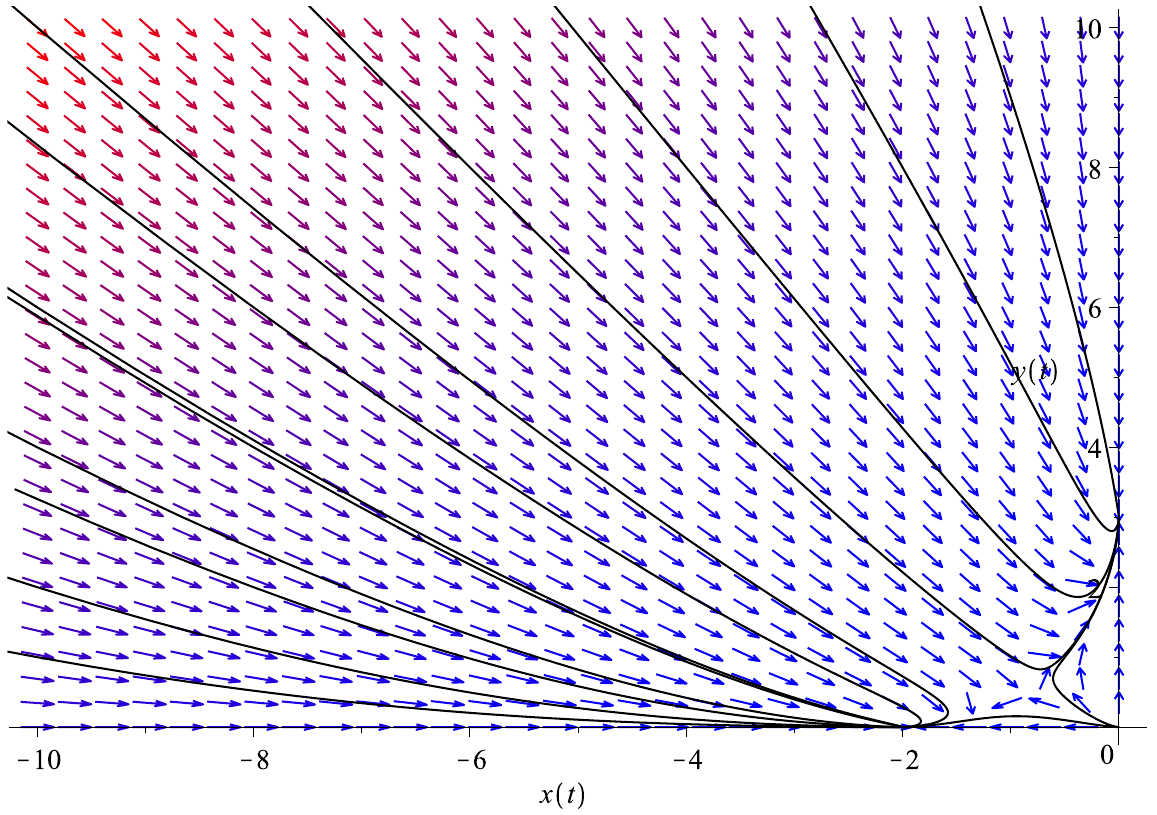}
\caption{The phase diagram of the two dark energy fluid model in Eq.(\ref{model3}). Here we use $r_1=1$, $K_1=2$, and $r_2=1.5$, $K_2=3$. The system exhibits trivial fixed points at $(x,y)=(0,K_2)$ and $(x,y)=(K_1,0)$, but also notice the existence of a saddle point that ``separates'' the two families of flow.}
\label{compflow}
\end{center}
\end{figure}

In population biology, two species that are very similar (i.e., fulfilling the same ecological niche) will compete for the same resources. To model such a situation, we consider both species to have a positive ``birth rate'', so that in this sense the competition is fair. In place of Eq.(\ref{model2}), we have:
\begin{equation}\label{model3}
\begin{cases}
\dfrac{\d x}{\d t}=r_1 x\left(1+\dfrac{x}{K_1}\right)-r_1xy,\\
\\
\dfrac{\d y}{\d t}= r_2 xy + r_2y\left(1-\dfrac{y}{K_2}\right),
\end{cases}
\end{equation}
where now $r_1:=-3H(1+w_{\Lambda_2})>0$. Note the coefficient pre-multiplying $r_1$ is now 1 instead of $-1$. 
Since $x<0$, we consider the resource term to be $1+x/K_1$ instead of $1-x/K_1$, keeping $K_1 > 0$.  
Note again that ``competition'' means that the interaction is harmful for both species, so the interaction term is negative for both fluids ($r_2 x y<0$ because $x<0$). For a fair competition we also include the carrying capacities $K_1$ and $K_2$ for both fluids, with $\mathcal{O}(K_1)=\mathcal{O}(K_2)$ and both being positive. A typical phase diagram is shown in Fig.(\ref{compflow}). Two families of flows are observed: those that flow towards $K_1$ and those that flow towards $K_2$. We are concerned with the latter.

\begin{figure}
\begin{center}
\includegraphics[width=0.45\textwidth]{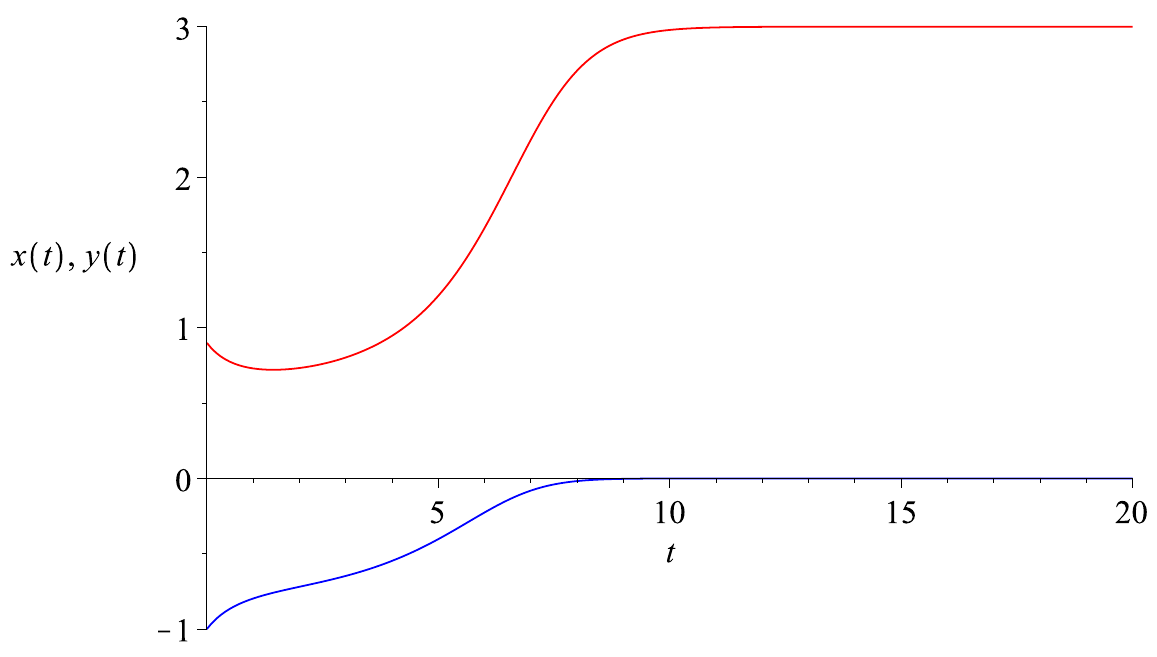}
\includegraphics[width=0.45\textwidth]{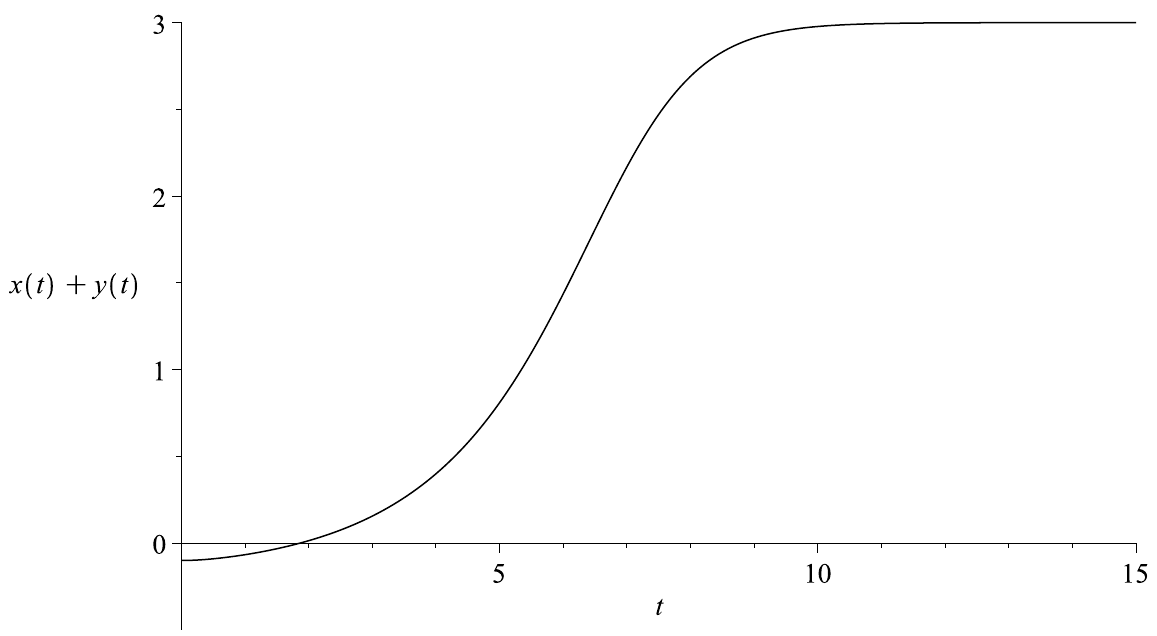}
\caption{\textbf{Top:} The evolution of $x(t)$ (bottom curve in blue) and $y(t)$ (top curve in red) with the initial condition set to be $x(0) = -1, y(0) = 0.9$, and with $r_1=1,r_2=1.5;K_1=2,K_2=3$. \textbf{Bottom:} The evolution of $x(t)+y(t)$, which is essentially the re-scaled version of  $\rho_{\Lambda_1}(t)+\rho_{\Lambda_2}(t)$.}
\label{ex2}
\end{center}
\end{figure}

\begin{figure}
\begin{center}
\includegraphics[width=0.46\textwidth]{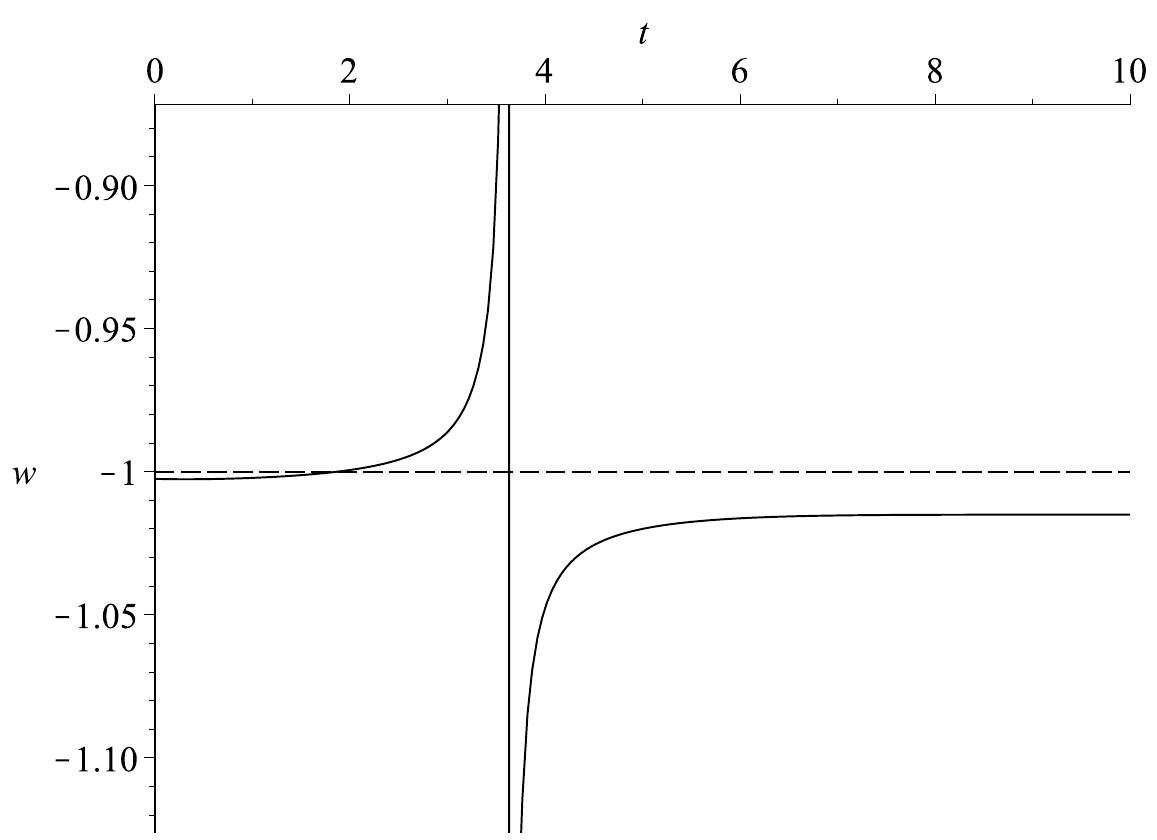}
\caption{The evolution of the overall effective $w$ of the example in Fig.(\ref{ex2}) with $w_{\Lambda_1}=-1.01$ and $w_{\Lambda_1}=-2.01$. There are two phantom crossings. The first crossing is smooth and corresponds to the time when $x+y=0$. The second crossing is singular and corresponds to $\rho_{\Lambda_1}+\rho_{\Lambda_2}=0$. These two conditions are not the same since $r_1 \neq r_2$.}
\label{compw}
\end{center}
\end{figure}

An explicit example of the rescaled energy density $x$ and $y$ as well as their sum are provided in Fig.(\ref{ex2}). 
Here we let $r_1=1$ and $r_2=1.5$.
We can see that $x(t)+y(t)$ changes sign in Fig.(\ref{compw}). Now, in this example, 
since $r_1 \neq r_2$ we cannot directly compare the sign change of $x(t)+y(t)$ to that of the overall dark energy density. However, we note that 
initially, $\text{sgn}(x+y)=\text{sgn}(\rho_{\Lambda_1}+1.5\rho_{\Lambda_2})=-1$. Since $\rho_{\Lambda_2}>0$, it follows that $\rho_{\Lambda_1}+\rho_{\Lambda_2}<\rho_{\Lambda_1}+1.5\rho_{\Lambda_2}<0$. 
Thus, it follows that $\text{sgn}(\rho_{\Lambda_1}+\rho_{\Lambda_2})=-1$ initially.
On the other hand, asymptotically we have $\text{sgn}(x+y) = \text{sgn}(y)=\text{sgn}({K_2})=+1$ as $x\to 0^-$. Equivalently, at late time $\text{sgn}(\rho_{\Lambda_1}+\rho_{\Lambda_2}) = \text{sgn}(\rho_{\Lambda_2})=+1$. Thus, we see that $\rho_{\Lambda_1}+\rho_{\Lambda_2}$ does indeed change sign. That is, the Universe transits from AdS-like to dS-like.

Incidentally, we also note that if $x(t)+y(t)$ is monotonically increasing, then we can give a bound on $\rho_{\Lambda_1}(t) + \rho_{\Lambda_2}(t)$. To see this, simply observe that
\begin{equation}
\frac{\text{d}(x(t)+y(t))}{\text{d}t} > 0
\end{equation}
is equivalent to
\begin{equation}
\frac{\gamma}{r_1r_2}(r_1\dot{\rho}_{\Lambda_1}+r_2\dot{\rho}_{\Lambda_2}) > 0.
\end{equation}
Given that $\gamma, r_1, r_2 > 0$, this means $r_1\dot{\rho}_{\Lambda_1}+r_2\dot{\rho}_{\Lambda_2}>0$.
Inserting a few terms that cancel each other yields:
\begin{equation}
r_1\dot{\rho}_{\Lambda_1}+(r_1\dot{\rho}_{\Lambda_2}-r_1\dot{\rho}_{\Lambda_2})+(r_2\dot{\rho}_{\Lambda_1}-r_2\dot{\rho}_{\Lambda_1})+r_2\dot{\rho}_{\Lambda_2}>0.
\end{equation}
Therefore, 
\begin{flalign}
(r_1+r_2)(\dot{\rho}_{\Lambda_1}+\dot{\rho}_{\Lambda_2}) &> r_2\dot{\rho}_{\Lambda_1}+r_1\dot{\rho}_{\Lambda_2} \notag \\
&=r_2^2\frac{\dot{x}}{\gamma}+r_1^2\frac{\dot{y}}{\gamma}.
\end{flalign}
If $r_2 > r_1$, we can write the last equation as 
\begin{equation}
\gamma^{-1}[r_1^2(\dot{x}+\dot{y})+(r_2^2-r_1^2)\dot{x}].
\end{equation}
Likewise, if $r_1>r_2$, we can write
\begin{equation}
\gamma^{-1}[r_2^2(\dot{x}+\dot{y})+(r_1^2-r_2^2)\dot{y}].
\end{equation}
Thus, for example, if $r_2>r_1$ (and hence $r_2^2>r_1^2$ -- recall that $r_1,r_2$ are positive), and we observe that $x+y$ and $x$ are both monotonicaly increasing, then so must ${\rho}_{\Lambda_1}+{\rho}_{\Lambda_2}$:
\begin{equation}
(r_1+r_2)(\dot{\rho}_{\Lambda_1}+\dot{\rho}_{\Lambda_2}) > \gamma^{-1}[r_1^2(\dot{x}+\dot{y})+(r_2^2-r_1^2)\dot{x}] > 0
\end{equation}

The phantom divide can be crossed as shown in Fig.(\ref{compw}), where $w$ is given by Eq.(\ref{w}). The values of $w_{\Lambda_1}$ and $w_{\Lambda_2}$ are constrained by the defining equations $r_1=-3H(1+w_{\Lambda_1})$ and  $r_2=-3H(1+w_{\Lambda_2})$. With $r_1=1$ and $r_2=1.5$, if we take $w_{\Lambda_1}=-1.01$, say, then $w_{\Lambda_1}=-1.015$. Note that there are \emph{two} phantom crossings\footnote{This phantom crossing is achieved by exhibiting a pole/singularity
in their equation of state parameter, which is quite different from the more well-known quintom models.} here: the first one occurs \emph{without} any singularity. From Eq.(\ref{w}) it can shown that if the denominator is not zero, then such a smooth crossing occurs precisely when $x+y=0$. This cannot happen for the unfair competition model as the condition would be $x-y=0$ instead (which cannot occur since $x<0$ and $y>0$). Note that despite the fact that there are two phantom crossings, there is only one AdS-to-dS transition in this example.

Even more surprising is the fact that the overall $w<-1$ can stay constant, yet there is still a AdS-to-dS transition. To achieve this we simply need to choose $r_1=r_2$. 
The plots of $x(t),y(t)$ and their sum is qualitatively the same as Fig.(\ref{ex2}) and are thus not shown. However, from the defining relations of $r_1$ and $r_2$ we would have $w_1=w_2$, and so in Eq.(\ref{w}), we obtain 
\begin{equation}
w=\frac{w_{\Lambda_1}\rho_{\Lambda_1}+w_{\Lambda_2}\rho_{\Lambda_2}}{\rho_{\Lambda_1}+\rho_{\Lambda_2}}=\frac{w_1(\rho_{\Lambda_1}+\rho_{\Lambda_2})}{\rho_{\Lambda_1}+\rho_{\Lambda_2}}=w_1.
\end{equation}
Strictly speaking during the transition point $\rho_{\Lambda_1}+\rho_{\Lambda_2}=0$, which otherwise would give rise to a singular behavior, we get an indeterminate form $0/0$, but both the left and right limit is well-defined and equal to $w_1$, so \emph{physically} it makes sense to say that $w\equiv w_1=w_2$ for all $t$. As such this evades the recent theorem that a sign-changing dark energy must have a singular equation of state \cite{2203.04167}. The reason this does not really violate the theorem therein is because the proof in \cite{2203.04167} is strictly for DE fluids that obey the usual continuity equation $\dot{\rho}+3H(\rho+p)=0$, whereas in our model it can be checked that the combined DE does not satisfy the continuity equation; the ``carrying capacity'' term $K$ breaks the continuity equation. This is equivalent to saying that $\nabla_\mu T_\text{DE}^{\mu\nu}\neq 0$.  This is not surprising -- as we will see in the Discussion section, our models have nontrivial nonlinear self-interaction term that acts as a source term for the continuity equation.

\section{The Conversion Model}\label{convert}

\begin{figure}[h!]
\begin{center}
\includegraphics[width=0.45\textwidth]{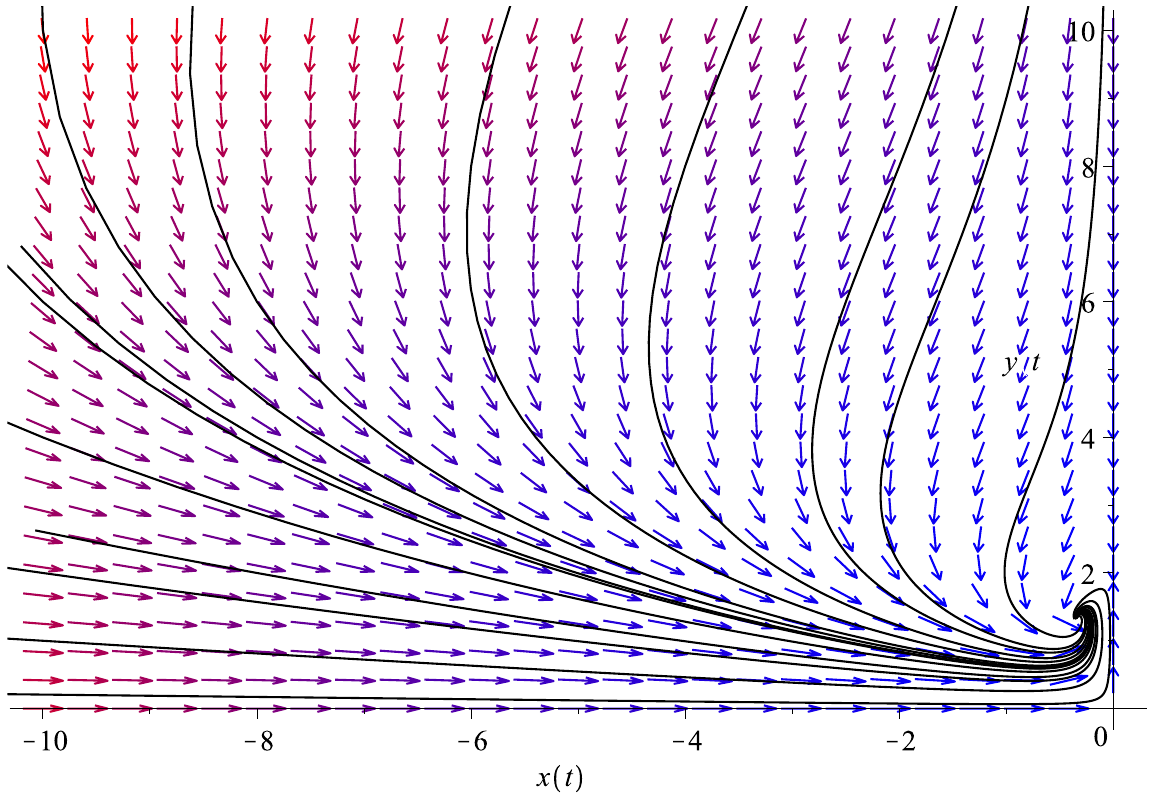}
\caption{The phase diagram of the two dark energy fluid model in Eq.(\ref{model4}). Here we use $r_1=1=r_2$, $K_1=1$, and $K_2=2$. The system exhibits an attractor fixed point at $(x,y)=(-1/3,3/2)$.}
\label{tcompwflow}
\end{center}
\end{figure}

Given the results above, one might wonder whether the AdS-to-dS transition can still happen if we restrict the growth of $\Lambda_2$ by converting it into $\Lambda_1$, or equivalently, by giving $\Lambda_1$ an advantage. This is achieved by the following model involving a quintessence $\Lambda_1$ and a phantom $\Lambda_2$:
\begin{equation}\label{model4}
\begin{cases}
\dfrac{\d x}{\d t}=-r_1 x\left(1-\dfrac{x}{K_1}\right)+r_1xy,\\
\\
\dfrac{\d y}{\d t}= r_2 xy + r_2y\left(1-\dfrac{y}{K_2}\right),
\end{cases}
\end{equation}
in which we note that the second term of $\d x/ \d t$ is now $+r_1 xy$ instead of $-r_1 xy$. The interaction is therefore beneficial to $x$ but harmful to $y$. This sounds like the complete opposite of what we wish to achieve (to have $\Lambda_2$ being the dominant term at late time). Surprisingly even in such a scenario it is possible to have a phantom crossing. The only difference being the attractor is now a stable spiral centered at 
\begin{equation}\label{spiral}
(x,y)=\left(-\frac{K_1(K_2-1)}{K_1K_2+1}, \frac{K_2(K_1+1)}{K_1K_2+1}\right),
\end{equation}
as can be seen in the example depicted in Fig.(\ref{tcompwflow}). 

\begin{figure}
\begin{center}
\includegraphics[width=0.45\textwidth]{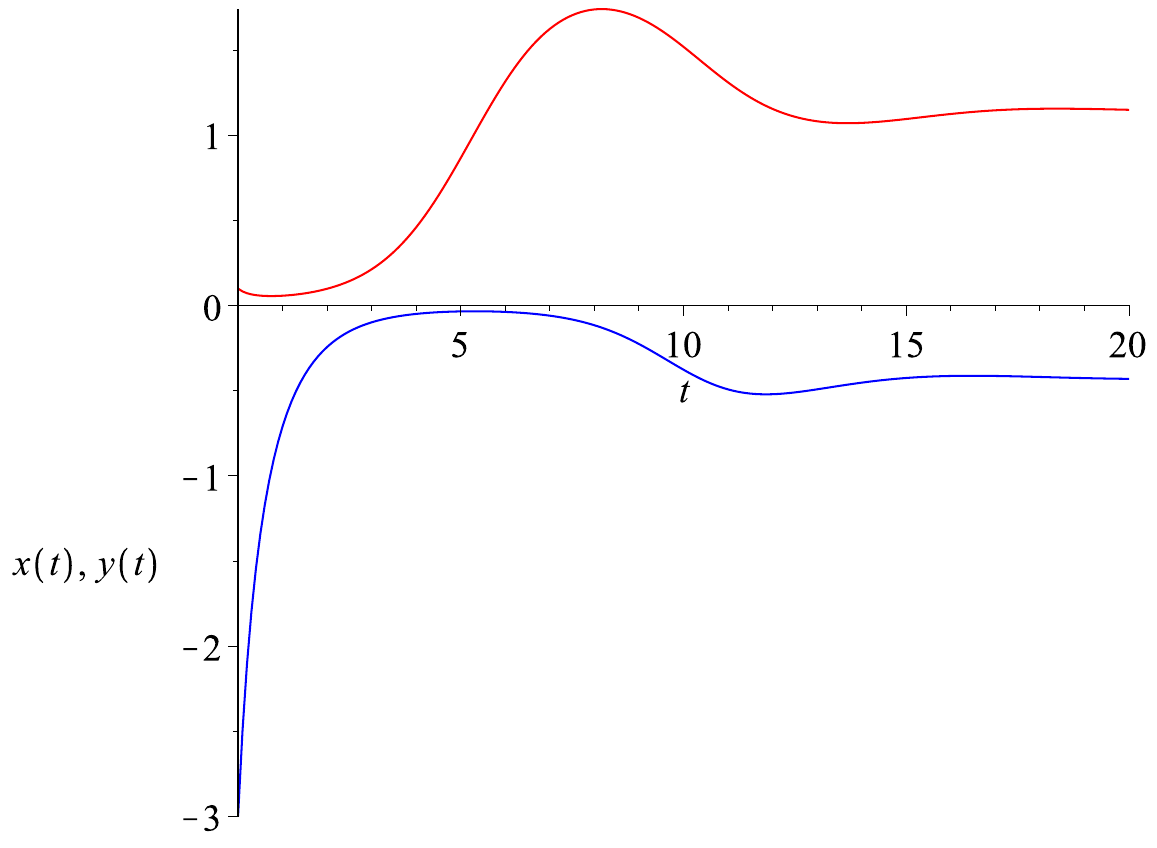}
\includegraphics[width=0.47\textwidth]{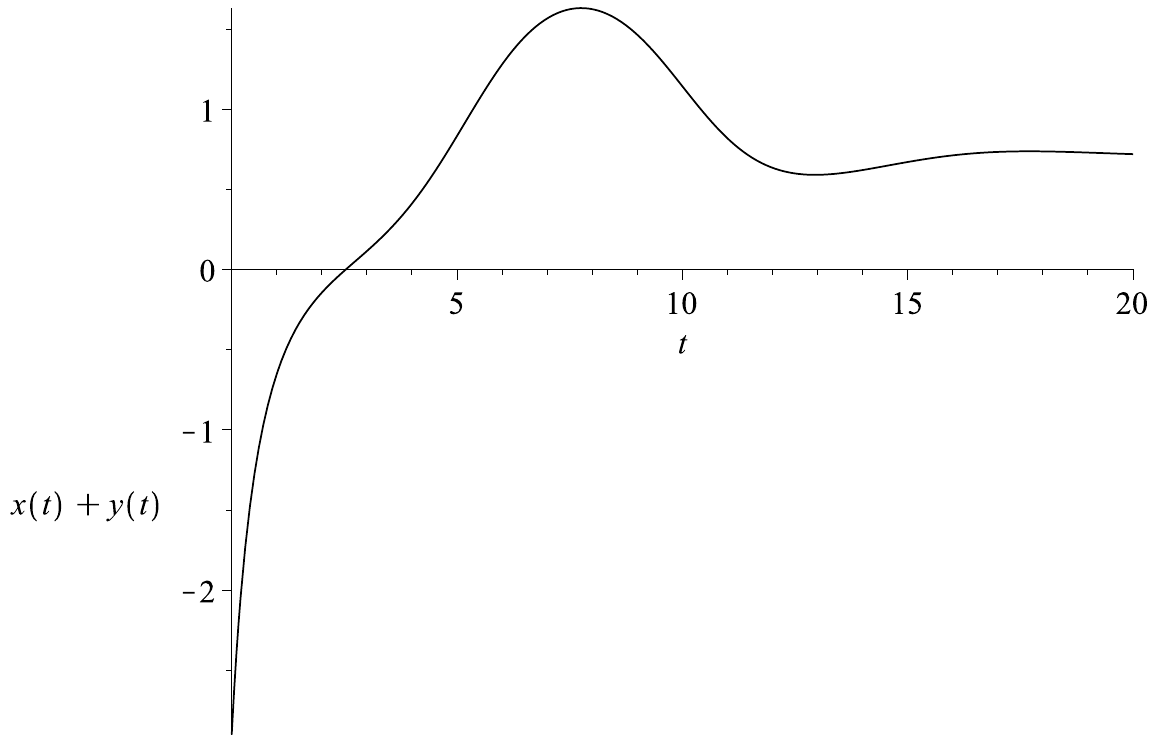}
\caption{\textbf{Top:} The evolution of $x(t)$ (bottom curve in blue) and $y(t)$ (top curve in red) with the initial condition set to be $x(0) = -3, y(0) = 0.1$, and with $r_1=1=r_2;K_1=1,K_2=2$. \textbf{Bottom:} The evolution of $x(t)+y(t)$, which in this case is the same as the overall $\rho$.}
\label{ex3}
\end{center}
\end{figure}

\begin{figure}
\begin{center}
\includegraphics[width=0.47\textwidth]{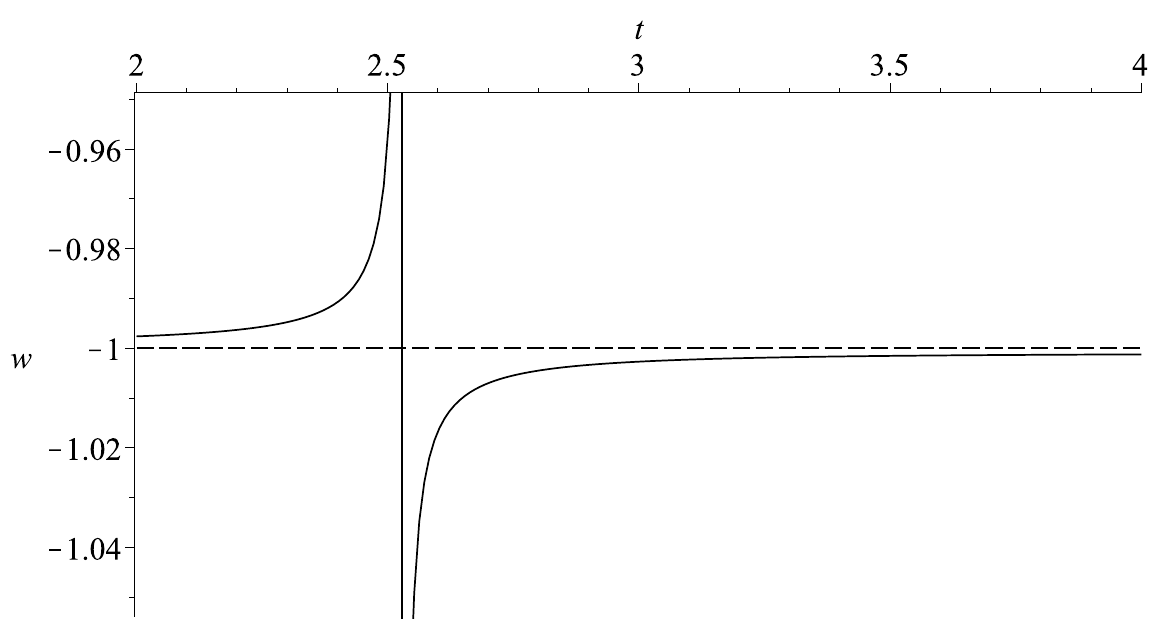}
\caption{The evolution of the overall effective $w$ of the example in Fig.(\ref{ex3}) with $w_{\Lambda_1}=-0.999$ and $w_{\Lambda_1}=-1.001$.}
\label{tcompw}
\end{center}
\end{figure}

What happens is that, despite the conversion term, $\Lambda_2$ can still dominate at late time. After all, we do not need $\Lambda_2$ to grow too big.
The evolution of $x(t)$ and $y(t)$ as well as their sum (which in this example is the same as $\rho_{\Lambda_1}+\rho_{\Lambda_2}$) are shown in Fig.(\ref{ex3}). The phantom crossing is shown in Fig.(\ref{tcompw}).

\begin{figure}[h!]
\begin{center}
\includegraphics[width=0.45\textwidth]{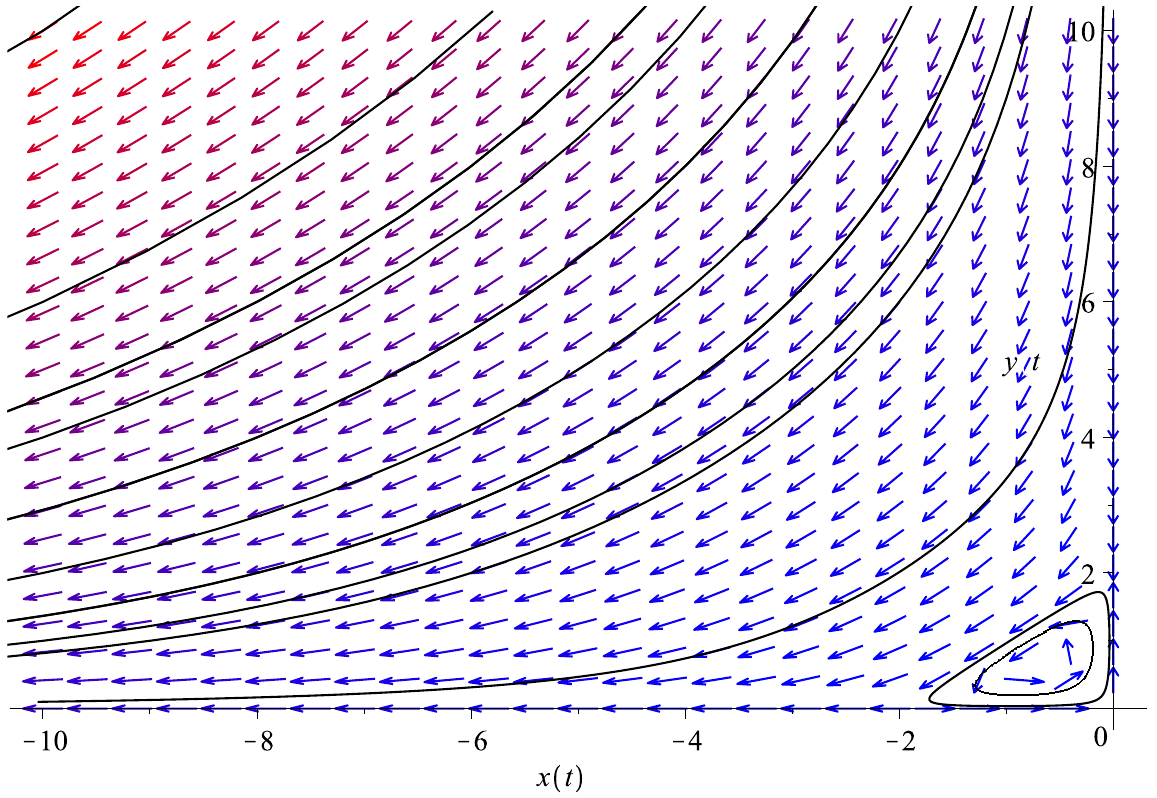}
\caption{The phase diagram of the two dark energy fluid model in Eq.(\ref{model4}) but with the resource term changed to $1-{x}/{K_1}$ instead of $1+{x}/{K_1}$. Here we use $r_1=1=r_2$, $K_1=2=K_2$. Most trajectories flow towards $x\to -\infty$ and $y\to 0$. Note, however, the presence of a center surrounded by cyclic flows.}
\label{cflow}
\end{center}
\end{figure}

The reason we consider a resource term $1-{x}/{K_1}$ instead of $1+{x}/{K_1}$ as in the previous section (incidentally, this puts the fixed point $x=K_1>0$ outside the physical phase space) is that
otherwise, with $1+{x}/{K_1}$ and the coefficient pre-multiplying $r_1$ being negative (quintessence) instead of positive (phantom), we observe that for ``most'' initial conditions,
\begin{equation}
\frac{\d |x|}{\d t} \sim \frac{r_1}{K_1}\frac{|x|^2}{2},
\end{equation}
and thus the magnitude of $x$ is increasing and $x \to -\infty$ instead of 0, which is not the behavior that we want. This can be seen in Fig.(\ref{cflow}).

\begin{figure}
\begin{center}
\includegraphics[width=0.45\textwidth]{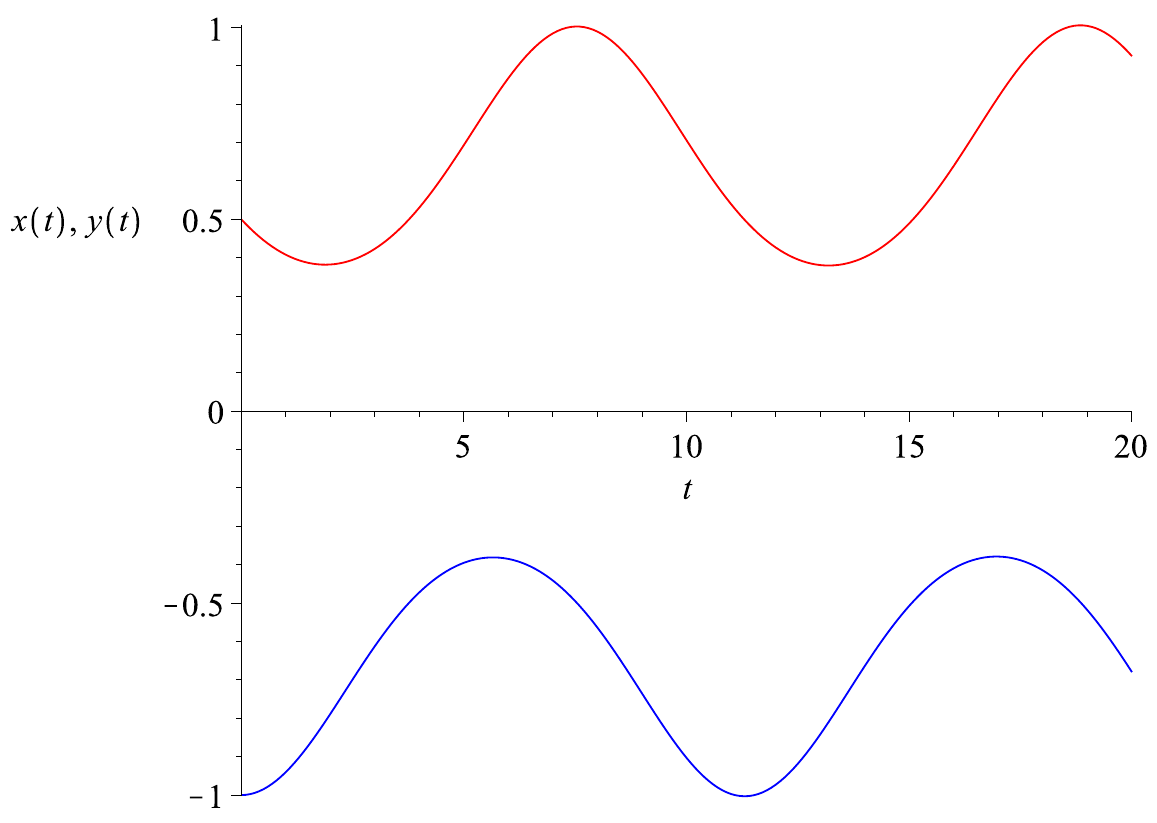}
\includegraphics[width=0.46\textwidth]{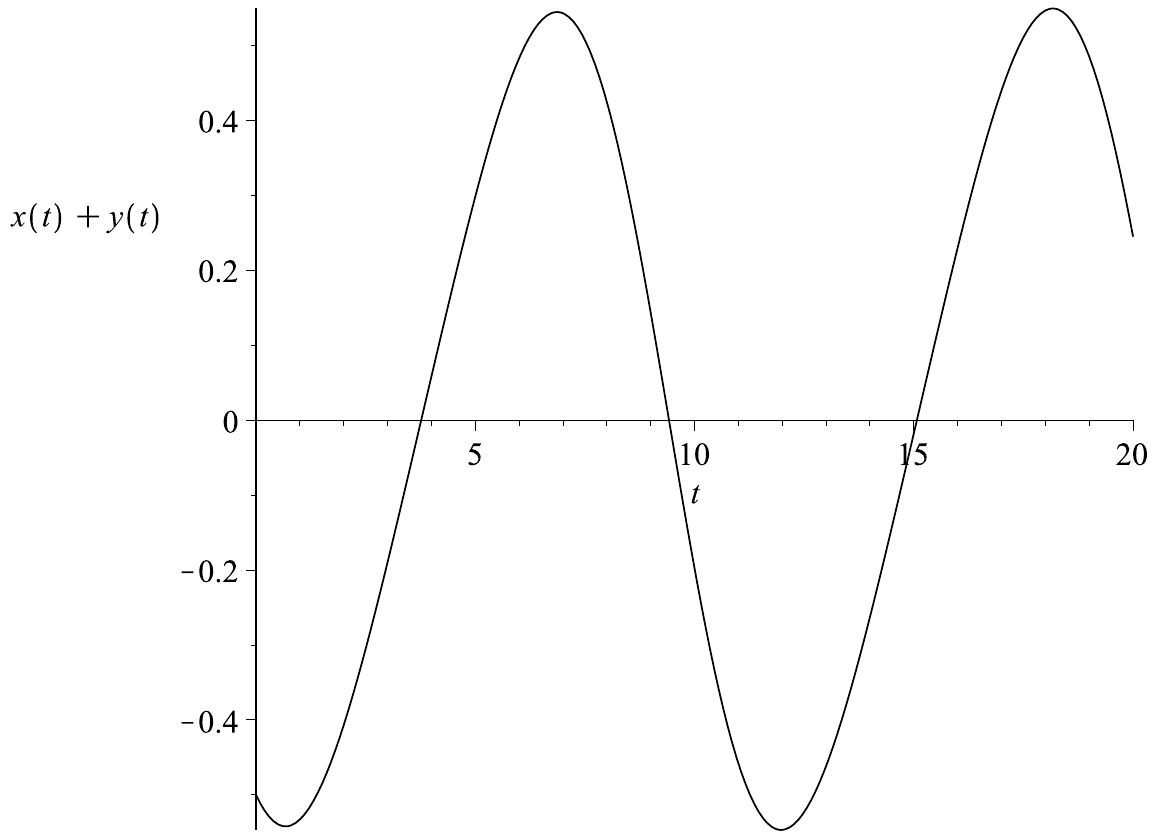}
\caption{\textbf{Top:} The evolution of $x(t)$ (bottom curve in blue) and $y(t)$ (top curve in red) with the initial condition set to be $x(0) = -1, y(0) = 0.5$, for the model that corresponds to Fig.(\ref{cflow}). \textbf{Bottom:} The evolution of $x(t)+y(t)$, which in this case is the same as the overall $\rho$.}
\label{ex4}
\end{center}
\end{figure}

\begin{figure}
\begin{center}
\includegraphics[width=0.46\textwidth]{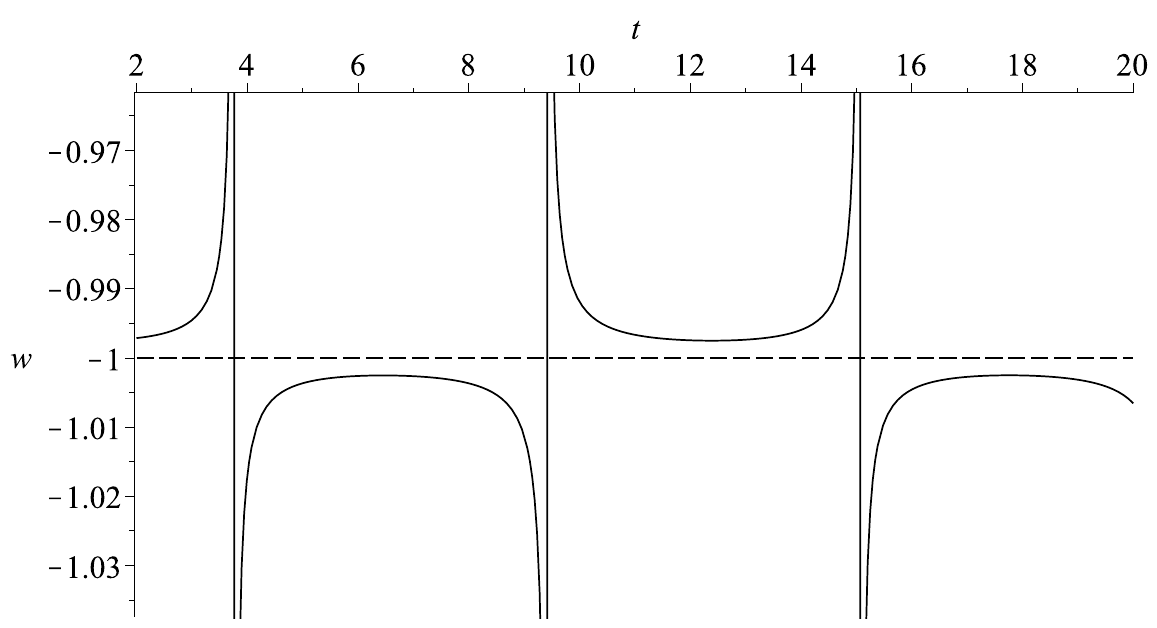}
\caption{The evolution of the overall effective $w$ of the example in Fig.(\ref{ex4}) with $w_{\Lambda_1}=-0.999$ and $w_{\Lambda_1}=-1.001$.}
\label{cw}
\end{center}
\end{figure}

However, even in this scenario there is one interesting feature worth mentioning. In the neighborhood of the origin, there exists a center around which the flows are cyclic. This implies both $x$ and $y$, as well as their sum, are oscillatory. As a result, there are multiple (infinitely many) phantom crossings, and infinitely many transitions between AdS-like to dS-like cosmology. 
These are shown in Fig.(\ref{ex4}) and Fig.(\ref{cw}). Indeed, multiple transition scenarios have been considered in the literature \cite{2208.05583}. 

\section{Discussion: Sign Switching Dark Energy and Naturalness}

One of the longstanding questions about dark energy density is why its value is so small, which is some $10^{-120}$ times smaller than the natural scale for a quantum vacuum energy if it is indeed a cosmological constant (for a dynamical field, the problem translates into an extremely light mass of the field). Of course, it is debatable whether this is indeed a problem \cite{1002.3966}. In any case, it would be interesting to see what this value corresponds to in the Lotka-Volterra equations in these models.

Take for example, the unfair competition model. We note that the evolution equation for $y(t)$, namely
\begin{equation}
\frac{\d y}{\d t}= r_2 xy + r_2y\left(1-\frac{y}{K}\right),
\end{equation}
is equivalent to the following fluid equation:
\begin{equation}\label{fluid}
\frac{\d \rho_{\Lambda_2}}{\d t} + 3H(\rho_{\Lambda_2}+p_{\Lambda_2})=\gamma \rho_{\Lambda_1}\rho_{\Lambda_2} + \frac{\gamma}{K}\left(\frac{1+w_{\Lambda_2}}{1+w_{\Lambda_1}}\right)\rho_{\Lambda_2}^2. 
\end{equation}
In other words, the ``resource term'' in the Lotka-Volterra equation corresponds to a quadratic self-interaction term. How might one interpret this term?

Such a term was also considered in \cite{1610.07338} and \cite{0512224}. As commented therein, pressure and density may not be linearly related in more complicated and more realistic systems. If we assume $p=p(\rho)$ for any barotropic fluid to be an analytic function, we can consider equation of state of the form $p=p_0+A_1\rho + A_2 \rho^2 + \mathcal{O}(\rho^3)$. This is a Taylor expansion of $p=p(\rho)$ about $\rho=0$, or upon re-grouping of terms, the expansion about the present energy density \cite{0512224,0309109}. 
If this is the correct interpretation, then the self-interaction term in Eq.(\ref{fluid}) can be interpreted as the result of the first order non-linear term in the expansion. 
However, in a series expansion, typically the \emph{coefficients} of the subsequent terms are roughly of the same order of magnitude, so the ``natural'' expectation is that 
 $\frac{\gamma}{K}\left(\frac{1+w_{\Lambda_2}}{1+w_{\Lambda_1}}\right) \sim \mathcal{O}(\gamma)$.
Even if $w_{\Lambda_1}$ and $w_{\Lambda_2}$ can be very close to $-1$, generically we would have the ratio $(1+w_{\Lambda_2} )/(1+w_{\Lambda_1})$ to be of order 1.  
This means that $K$ being small $\mathcal{O}(\varepsilon) \ll 1$ would typically result in the quadratic coefficient being unnaturally large and dominate over the linear term, which in turn suggests that we should \emph{not}, in fact, interpret this term as a term in a Taylor series expansion of $p_{\Lambda_2}=p_{\Lambda_2}(\rho_{\Lambda_2})$. Note that if we do not interpret $\rho_{\Lambda_2}$ and $\rho_{\Lambda_2}^2$ term as part of a Taylor series, we \emph{can} still absorb the $\rho_{\Lambda_2}^2$ term as part of the pressure so that $p_{\Lambda_2} = w_2 \rho_{\Lambda_2} + \text{const.} \rho_{\Lambda_2}^2$. In which case $K$ is related to the mass scale $M$ of $\Lambda_2$ via $K \sim M^4$, see \cite{0507120,1301.4746}; so this is just the aforementioned fact that in the case of dark energy being dynamical, the naturalness problem is its small mass scale. 
In the conversion model, the situation is similar. The attractor of the spiral is given in Eq.(\ref{spiral}). In which we see that $y$ and $x$ are both small if $K_1$ and $K_2$ are small.
Obviously, our models do not solve the naturalness problem, unless one could explain dynamically why the attractor has such a small value.
Perhaps a fundamental understanding of the nature of the phantom fluid or an entropic argument could provide such a mechanism. (In the context of cosmological constant, it was argued in \cite{2210.01142} that gravitational entropy is maximized by $\Lambda \to 0^+$.)

To conclude, in this work, motivated by the idea that a sign switching dark energy from an early time AdS-like Universe to a late time dS-like Universe can help to ameliorate the Hubble tension and the $S_8$ tension, we consider a scenario in which the dark energy sector consists of two interacting fluids. We found that AdS-to-dS transition can happen under various models, even if both fluids are phantom, and even if the combined effective dark energy has a constant $w$. Of course, these are only toy models serve to illustrate the qualitative features. The profiles of these fluids need to be constrained by observations. Still, the possibility that the dark energy sector may contain various interacting components deserves a closer look (see \cite{1301.4746,1705.04737} for other aspects of self-interacting dark energy) as it can realize many different interesting features.

For generalizations, one could also consider interactions between the two dark energy components with dark matter and/or dark radiation in a more complicated model (a quintom model was considered in \cite{1908.03324}, with the phantom component interacting with dark matter, but not with the quintessence sector; see also \cite{0805.2255}). Another possibility is to consider other forms of interaction terms in place of $\pm\gamma \rho_{\Lambda_1}\rho_{\Lambda_2}$; see for example \cite{2103.13432} and the references therein for the case of DM-DE interaction. Most importantly, more realistic models need to go beyond the assumption that the Hubble parameter is slowly varying when setting up the Lotka-Volterra equations.

\begin{acknowledgments}
YCO thanks the National Natural Science Foundation of China (No.11922508) for funding support. He also thanks Brett McInnes for fruitful discussions, and an anonymous reviewer for useful comments.
\end{acknowledgments}

\end{document}